\documentstyle[12pt]{article}

\newcommand{\eq}{\begin{equation}}
\newcommand{\eqa}{\begin{eqnarray}}
\newcommand{\en}{\end{equation}}
\newcommand{\ena}{\end{eqnarray}}
\newcommand{\enn}{\nonumber \end{equation}}

\def\sk{\vskip .4cm}
\def\noi{\noindent}
\def\om{\omega}

\def\Ga{\Gamma}
\def\de{\delta}

\def\la{q-q^{-1}}
\def\lam{{1 \over \la}}

\def\thetap{{\theta}^{\prime}}

\def\unmezzo{{1 \over 2}}
\def\epsi{\varepsilon}
\def\we{\wedge}

\def\de{\delta}

\def\part{\partial}

\def\R#1#2{ R^{#1}_{~~~#2} }

\def\Rinv#1#2{ (R^{-1})^{#1}_{~~~#2} }

\def\Rhat#1#2{ \Rh^{#1}_{~~~#2} }
\def\Rhatinv#1#2{ (\Rh^{-1})^{#1}_{~~~#2} }
\def\Z#1#2{ Z^{#1}_{~~~#2} }

\def\Rh{\Lambda}

\def\MM#1#2#3{M^{#1~~~#3}_{~#2}}
\def\cchi#1#2{\chi^{#1}_{~#2}}
\def\ome#1#2{\om_{#1}^{~#2}}
\def\RRhat#1#2#3#4#5#6#7#8{\Rh^{~#2~#4}_{#1~#3}|^{#5~#7}_{~#6~#8}}
\def\RRhatinv#1#2#3#4#5#6#7#8{(\Rh^{-1})^
{~#2~#4}_{#1~#3}|^{#5~#7}_{~#6~#8}}

\def\Cb{{\bf C}}
\def\CC#1#2#3#4#5#6{\Cb_{~#2~#4}^{#1~#3}|_{#5}^{~#6}}
\def\cc#1#2#3#4#5#6{C_{~#2~#4}^{#1~#3}|_{#5}^{~#6}}

\def\DR{\Delta_R}
\def\DL{\Delta_L}

\def\T#1#2{ T^{#1}_{~~#2} }
\def\Ti#1#2{ (T^{-1})^{#1}_{~~#2} }

\def\qm{q^{-1}}

\def\D{\Delta}

\def\Ap{A^{\prime}}

\def\kpr{k^{\prime}}

\def\qone{$q \rightarrow 1~$}

\def\ra{\rightarrow}
\def\detq{{\det}_q}

\begin{document}

\begin{titlepage}
\rightline{CERN--TH./92}
\rightline{DFTT-44/92}
\vskip 2em
\begin{center}{\bf BICOVARIANT DIFFERENTIAL GEOMETRY}\\
{\bf OF THE QUANTUM GROUP $GL_q(3)$}
\\[6em]
 Paolo Aschieri${}^{\diamond *}$
and Leonardo Castellani${}^{*}$ \\[2em]
{\sl${}^{\diamond}$ CERN, CH-1211 Geneva 23, Switzerland.\\
{}$^{*}$Istituto Nazionale di
Fisica Nucleare, Sezione di Torino\\
and\\Dipartimento di Fisica Teorica\\
Via P. Giuria 1, 10125 Torino, Italy.}  \\[6em]
\end{center}
\begin{abstract}
We construct a bicovariant differential calculus on the quantum
group $GL_q(3)$, and discuss its restriction to $[SU(3) \otimes U(1)]_q$.
The $q$-algebra of Lie derivatives is found, as well as the Cartan-Maurer
equations. All the quantities characterizing the non-commutative
geometry of $GL_q(3)$ are given explicitly.

\end{abstract}

\vskip 4cm

\noi CERN--TH./92

\noi DFTT-44/92

\noi July1992
\noi \hrule
\vskip.2cm
\hbox{\vbox{\hbox{{\small{\it email addresses:}}}\hbox{}}
 \vbox{\hbox{{\small Decnet=(39163::ASCHIERI, CASTELLANI;
VXCERN::ASCHIERI)}}
\hbox{{\small Bitnet=(ASCHIERI, CASTELLANI@TORINO.INFN.IT )}}}}

\end{titlepage}
\newpage
\setcounter{page}{1}


It is tempting to use quantum groups \cite{Drinfeld}--\cite{Majid1} for
the construction of generalized gauge and gravity theories.
Some initial work in this direction can be found
in refs. \cite{Isaev}-\cite{Brz}
for $q$-gauge theories and in \cite{Cas1}, \cite{Aschieri}
for $q$-gravity theories.
The basic idea of ref. \cite{Cas1} is
to obtain a $q$-gravity theory as a ``gauge''
theory of the quantum Poincar\'e
group (or of the quantum De-Sitter group).
For this we need to study the
differential (non-commutative) geometry
of the relevant $q$-Poincar\'e (or $q$-De Sitter)
group. As a first step, the bicovariant
calculus on the $D=2$ $q$-Poincar\'e group was developed in \cite{Cas1}.
\sk
On the other hand, the differential
geometry of the quantum $SU(2)$ and $SU(3)$ groups
is of interest in the construction of $q$-deformed gauge
theories that reduce to the standard model
for $q\ra 1$. The Cartan-Maurer equations for
the $q$-analogue of the left-invariant one-forms
{\sl define} the field strengths, as in
the classical case,  and one can find an action of the
type $F^i_{\mu\nu} F^j_{\mu\nu} g_{ij}$ invariant
under the $q$-group symmetries \cite
{Cas2}.
\sk
The differential calculus on $q$-groups,
initiated by Woronowicz \cite{Wor}, has
been developed in a number of papers from various points of
view \cite{Manin}-\cite{Schupp}. The general
theory being still in its infancy,
we feel that explicit non-trivial examples may be inspiring, and
in this Letter we present a (bicovariant)
differential calculus on $GL_q(3)$ and on its restriction
to $[SU(3) \otimes U(1)]_q$.
\sk
The quantum group $GL_q(3)$ is defined \cite{FRT} as the (associative)
algebra $A$ freely generated by:
\sk
{\bf i)} non-commuting matrix entries
$\T{a}{b}$, {\small a,b=1,2,3}, satisfying
\eq
\R{ab}{ef} \T{e}{c} \T{f}{d} = \T{b}{f} \T{a}{e} \R{ef}{cd} \label{RTT}
\en
\noi where the $R$ matrix is given in Table 1, the rows and columns being
numbered in the order $11,12,13,21,22,23,31,32,33$. The $R$ matrix
satisfies the quantum Yang-Baxter equation
\eq
\R{a_1b_1}{a_2b_2} \R{a_2c_1}{a_3c_2} \R{b_2c_2}{b_3c_3}=
\R{b_1c_1}{b_2c_2} \R{a_1c_2}{a_2c_3} \R{a_2b_2}{a_3b_3}, \label{QYB}
\en
\noi a sufficient condition for the
consistency of the ``RTT'' relations (\ref{RTT}). One
can also think of the indices ${}^a_{~b}$ of
$\T{a}{b}$ as composite indices taking
the values $1,...9 $ (with $11\!\ra \!1,~ 12\!\ra\!2,
{}~ 13\!\ra \!3,~ 21\!\ra \!4,~
22\!\ra \!5,~
23\!\ra \!6,~ 31\!\ra \!7,~
 32\!\ra \!8,~ 33\!\ra \!9$), so that
\eq
\T{a}{b} = \left( \begin{array}{ccc} T_1&T_2&T_3\\
                                  T_4&T_5&T_6\\
                             T_7&T_8&T_9   \end{array} \right)
\en
Then the relations (\ref{RTT}) take the explicit form:
\eq
\begin{array}{lll}
T_1 T_2=qT_2 T_1,      &T_1 T_3=qT_3T_1 , &T_2T_3=qT_3T_2  ,      \\
T_1 T_4=qT_4 T_1,     &T_1 T_5=(q-\qm)T_4T_2+T_5T_1 ,
&T_1T_6=(q-\qm)T_4T_3
+T_6T_1,\\
T_2 T_5=qT_5 T_2,     &T_2T_6=(q-\qm)T_5T_3+T_6T_2 ,
&T_1T_9=(q-\qm)T_7T_3+T_9T_1,\\
T_2T_8=qT_8T_2  ,  &T_2T_9=(q-\qm)T_8T_3+T_9T_2,
&T_3T_9=qT_9T_3,\\
T_4T_5=qT_5T_4 ,&T_4T_6=qT_6T_4 , &T_5T_6=qT_6T_5,\\
T_4T_7=qT_7T_4 , &T_4T_8=(q-\qm)T_7T_5+T_8T_4 ,&T_4T_9=(q-\qm)T_7T_6
+T_9T_4,\\
T_5T_8=qT_8T_5, &T_5T_9=(q-\qm)T_8T_6+T_9T_5, &T_6T_9=qT_9T_6,\\
T_7T_8=qT_8T_7, &T_7T_9=qT_9T_7, &T_8T_9=qT_9T_8.
\end{array}
\label{RTTexplicit}
\en
\noi all other commutations
being trivial.  For $q \ra 1$ all elements
commute (the $\R{ab}{cd}$ matrix becomes $\de^a_c \de^b_d$, cf. Table 1).
\sk
{\bf ii)} the identity element $I$ and the inverse $\xi$ of
the $q$-determinant of $T$, defined by:
\eq \xi \detq T=\detq T\xi=I\label{xi} \en
\eq \detq T \equiv \sum_{\sigma} (-q)^{l(\sigma)} \T{1}{\sigma(1)} \cdots
\T{n}{\sigma(n)} \label{qdet} \en
\noi where $l(\sigma)$ is the minimum number of
transpositions in the permutation
$\sigma$. Explicitly:
\eq
\detq T=T_1T_5T_9+q^2 T_2T_6T_7+q^2 T_3T_4T_8-qT_1T_6T_8-q^3T_3T_5T_7-
qT_2T_4T_9 ~. \label{detT}\en
\sk
\noi It is not difficult to
check that $\xi$ and $\detq T$ commute with all
the elements $\T{a}{b}$. The non-commutative
algebra generated by $\T{a}{b}$, $I$ and
$\xi$ is a Hopf algebra, i.e.
we can define a coproduct $\D$, a
counit $\epsi$ and a coinverse $\kappa$:
\eqa
& &\D(\T{a}{b})=\T{a}{c}
\otimes \T{c}{b},~~\D (I)=I\otimes I\label{coproductT}\\
& & \epsi(\T{a}{b})=\de^a_b,~~\epsi (I)=1\label{couT}\\
& & \kappa(\T{a}{b})=\Ti{a}{b},~~\kappa(I)=I. \label{coiT}
\ena
\noi the (matrix) inverse of $T$ being given by(cf. \cite{FRT}):
\eq
\Ti{a}{b}= \xi \left( \begin{array}{ccc}
         T_5T_9-qT_6T_8&-\qm T_2T_9+T_3T_8&q^{-2}T_2T_6-\qm T_3T_5\\
         -q T_4T_9+q^2 T_6T_7&T_1T_9-qT_3T_7&-\qm T_1T_6+T_3T_4\\
         q^2T_4T_8-q^3T_5T_7&-qT_1T_8+q^2T_2T_7&T_1T_5-qT_2T_4\\
                           \end{array} \right)\nonumber
\en
\noi From eqs. (\ref{detT}) and (\ref{xi}) one
deduces the co-structures on $\detq T$ and
$\xi$:
\eqa
& &  \D (\detq T)=\detq T \otimes \detq T, ~~\D (\xi)=\xi \otimes \xi\\
& &  \epsi (\detq T)=1,~~\epsi (\xi)=1\\
& &  \kappa (\detq T)=\xi,~~\kappa (\xi)=\detq T
\ena
For $q \in R$ a $q$-analogue of complex conjugation ($*$-structure)
can be consistently defined on $A$ as \cite{FRT}:
\eq
T^*=(T^{-1})^t \label{unitarity}
\en
\noi This unitarity condition restricts $GL_q(3)$ to its
quantum subgroup
$[SU(3) \otimes U(1)]_q$. The further constraint
$\detq T=I$ yields $SU_q(3)$. Other real forms of $GL_q(3)$ can be
considered, via a more general unitarity condition \cite{FRT}.
\sk
The differential calculus we are about to construct can
be entirely formulated in terms of the $R$ matrix. This is true for
all quantum groups of the $A,B,C,D$ series. The
general constructive procedure can be
found in  ref. \cite{Jurco}, or, in the
notations we adopt here, in ref. \cite{Aschieri}.
There the reader can find the derivation of the general formulas
appearing later in this Letter.
\sk
As discussed in \cite{Wor} and \cite{Jurco},
we can start by introducing
the (quantum) left-invariant one-forms $\ome{a}{b}$, whose
exterior product
\eq
\ome{a_1}{a_2} \we \ome{d_1}{d_2}
\equiv \ome{a_1}{a_2} \otimes \ome{d_1}{d_2}
- \RRhat{a_1}{a_2}{d_1}{d_2}{c_1}{c_2}{b_1}{b_2}
\ome{c_1}{c_2} \otimes \ome{b_1}{b_2} \label{exteriorproduct}
\en
\noi is defined by the braiding matrix $\Rh$:
\eq
\RRhat{a_1}{a_2}{d_1}{d_2}{c_1}{c_2}{b_1}{b_2}
\equiv  d^{f_2} d^{-1}_{c_2} \R{f_2b_1}{c_2g_1} \Rinv{c_1g_1}{e_1a_1}
    \Rinv{a_2e_1}{g_2d_1} \R{g_2d_2}{b_2f_2} \label{RRffMM}
\en
\noi For $q\ra 1$ the braiding
matrix $\Lambda$ becomes the usual permutation
operator and one recovers the classical exterior product.

In the case of $GL_q(3)$ there are
nine independent left-invariant
one-forms $\ome{a}{b}$, {\small a,b=1,2,3}, and
we number them $\om^i$, {\small i=1,...9}, according
to the same index convention used for
the matrix elements $\T{a}{b}$, i.e. $\ome{1}{1} \ra
\om^1, \ome{1}{2} \ra \om^2$ etc. The $d$
vector in (\ref{RRffMM}) is given by
$d_1=q,d_2=q^3,d_3=q^5$.
\sk
The commutation relations for exterior products of $\ome{a}{b}$
are deduced
from the formula [valid for all $GL_q(N)$], cf. \cite{Aschieri}:
\eq
\om^i \we \om^j = - \Z{ij}{kl} \om^k \we \om^l \label{commom}
\en
\eq
\Z{ij}{kl} \equiv {1\over {q^2 + q^{-2}}} [\Rhat{ij}{kl} + \Rhatinv
{ij}{kl}]. \label{defZ}
\en
For $GL_q(3)$, these relations
are collected in Table 2 (we have omitted the
wedge symbol in the exterior products between the $\om$).
\sk
As in the classical case, the left-invariant
1-forms $\ome{a}{b}$ are a basis for the
space $\Gamma$ of quantum 1-forms, i.e. any $\rho \in \Gamma$
can be expressed as
 $\rho=a_b{}^a\,\ome{a}{b}$, with $a_b{}^a \in A$.

The left and right action of the
quantum group on $\Ga$ are defined as follows:
\eq
\DL (\ome{a_1}{a_2})=I\otimes \ome{a_1}{a_2},\label{DL}
\en
\eq
\DR (\ome{a_1}{a_2}) = \ome{b_1}{b_2} \otimes \MM{b_1}{b_2a_1}{a_2},
\en
\noi where $\MM{b_1}{b_2a_1}{a_2}$, the adjoint representation,
is given by
\eq
\MM{b_1}{b_2a_1}{a_2} \equiv \T{b_1}{a_1} \kappa (\T{a_2}{b_2}).
\label{adjoint}
\en
\noi The $q$-analogue of the fact
that left and right actions commute is given by the property
\eq
(id \otimes \DR)\DL=(\DL \otimes id)\DR
\en

The bimodule \footnote{``bimodule" meaning
that $\rho \in \Ga$ can be multiplied
on the right and on the left
by elements $a \in A$. In general $a\rho \not= \rho a$.}
$\Ga$ is further characterized by the commutations
between $\ome{a}{b}$ and the basic elements of $A$:
\eq
\ome{a_1}{a_2}\T{b_1}{b_2}=s\,\T{b_1}{d}
\Rinv{dc_1}{ea_1}\Rinv{a_2e}{c_2b_2}
\ome{c_1}{c_2} \label{omT}
\en
\eq
\ome{a_1}{a_2}\detq T=s^3 q^{-2}~ \detq T \ome{a_1}{a_2}~~,~~~~
\ome{a_1}{a_2}\xi= s^{-3} q^2 ~\xi
\ome{a_1}{a_2} \label{omdetT}
\en
\noi where $s$ is a complex arbitrary scale
(whose classical limit is $1$),
cf. refs. \cite{Sun} and \cite{Aschieri}. From eq. (\ref{omdetT})
we see that $s=q^{{2\over 3}}$ is a ``canonical" choice for $s$ (then
$\detq T$ and its inverse $\xi$ commute with any element of $\Ga$).
Equations (\ref{omT})
also allow to find the commutations between a
generic $\rho \in \Ga$ and  a generic $a \in A$, since
any $\rho$ is a linear combination
of $\ome{a}{b}$ with coefficients belonging to $A$, and any  $a$
is a polynomial in the $\T{a}{b}$.

Table 3 contains a list of the commutations (\ref{omT}) for $GL_q(3)$,
with $s=1$. The case $s \not= 1$ is recovered by multiplying
the right-hand sides of Table 3 by $s$.
\sk
The induced $*$-structure on the $\om$'s is given by \cite{Watamura},
\cite{Jurco}:
\eq
(\ome{a_1}{a_2})^*=-\ome{a_2}{a_1}
\en
\noi and is extended to $\Ga$ via the rules
\eq
(a\rho)^*=\rho^*a^*,~~~(\rho a)^*=a^* \rho^*.
\en
\noi This $*$-structure is compatible with the left and right
action of the quantum group, i.e.
\eq
\D (a^*)=[\D(a)]^*, ~~\forall a \in A
\en
\eq
\DL (\rho^*)=[\DL (\rho)]^*,~~\DR (\rho^*)=[\DR (\rho)]^*,~~\forall
\rho \in \Ga
\en
\noi where $(a \otimes b)^*\equiv b^* \otimes a^*,~(a\otimes \rho)^*
\equiv
a^* \otimes \rho^*,~(\rho \otimes a)^*\equiv\rho^* \otimes a^*,~~
\forall a,b \in A,~\rho \in \Ga $.
\sk
The exterior differential
on $\Ga^{\we k}=\Ga \we \Ga \we \cdots \we  \Ga$
(k-times)
is defined by means of the
bi-invariant (i.e. left- and right-invariant) element $\tau=\sum_a
\ome{a}{a} \in \Ga$ as follows:
\eq
d\theta \equiv \lam [\tau \we \theta - (-1)^k \theta \we \tau],
\label{exteriordifferential}
\en
\noi where $\theta \in \Ga^{\we k}$. The
normalization $\lam$ is necessary in order to obtain the
correct classical limit. For
$a \in A$ we have
\eq
da=\lam [\tau a - a \tau]. \label{exterior differentialonA}
\en
\noi This linear map satisfies the Leibniz rule
\eq
d(ab)=(da)b+a(db),~~\forall a,b\in A; \label{Leibniz}
\en
\noi and has the properties
\eq
d(\theta \we \thetap)=d\theta \we \thetap + (-1)^k \theta \we d\thetap
\label{propd1}
\en
\eq
d(d\theta)=0\label{propd2}
\en
\eq
d(\theta^*)=(d\theta)^*
\en
\eq
\DL (d\theta)=(id\otimes d)\DL(\theta)\label{propd3}
\en
\eq
\DR (d\theta)=(d\otimes id)\DR(\theta),\label{propd4}
\en
\noi where $\theta \in \Ga^{\we k}$, $\thetap \in \Ga^{\we \kpr}$.
The $*$-structure on $\Ga^{\we n}$ can be inferred from the rule
\eq
(\theta \we \thetap)^*=(-1)^{k\kpr}~{\thetap}^* \we \theta^*
\en
The properties (\ref{propd3}), (\ref{propd4}) express
the fact that $d$ commutes with the left
and right action of the quantum group, as in the classical case.
Moreover, for a generic $\rho \in \Ga$ we have
\eq
{}~~~~~~\rho=a_kdb_k ,~~~~~~~~~~~ a_k,b_k \in A \label{adb}
\en
The relations of Table 6 can be inverted to yield the
$\ome{a}{b}$ in terms of the $dT^c{}_d$, thus proving
the decomposition (\ref{adb}).
\sk
The bimodule $\Ga$ and the mappings $d, \DR, \DL$
with the above properties
define a {\sl bicovariant} differential
calculus on the $q$-group \cite{Wor}.
\sk
The ``quantum Lie algebra generators" are linear functionals on $A$
(i.e. belonging  to its dual $\Ap$)
introduced via the formula \cite{Wor}
\eq
da=\lam[\tau a - a\tau] =(\cchi{a_1}{a_2} * a)
\ome{a_1}{a_2}. \label{qgenerators}
\en
\noi where
\eq
{}~~\chi * a \equiv (id \otimes \chi) \D (a),~~~~\forall a \in A,~
\chi \in \Ap
\en
By taking the exterior derivative
of (\ref{qgenerators}), requiring that $d^2=0$ and
using the fact that $\tau=\ome{b}{b}$
is bi-invariant, we arrive at the $q$-Lie
algebra relations \cite{Jurco}, \cite{Aschieri}:
\eq
\cchi{d_1}{d_2} \cchi{c_1}{c_2} - \RRhat{e_1}{e_2}{f_1}{f_2}
{d_1}{d_2}{c_1}{c_2} ~\cchi{e_1}{e_2} \cchi{f_1}{f_2} =
\CC{d_1}{d_2}{c_1}{c_2}{a_1}{a_2} \cchi{a_1}{a_2}
\label{qLiealgebra}
\en
\noi where the structure constants are explicitly given by:
\eq
\CC{a_1}{a_2}{b_1}{b_2}{c_1}{c_2} =\lam [- \de^{b_1}_{b_2}
\de^{a_1}_{c_1}
\de^{c_2}_{a_2} + \RRhat{b}{b}{c_1}{c_2}{a_1}{a_2}{b_1}{b_2}]. \label{CC}
\en
\noi and $\cchi{d_1}{d_2}
\cchi{c_1}{c_2} \equiv (\cchi{d_1}{d_2}  \otimes
\cchi{c_1}{c_2}) \D$.

The $*$-structure on $A$ induces a $*$-operation on the $q$-generators
as follows:
\eq
(\cchi{a_1}{a_2})^*(a)=\overline{\cchi{a_2}{a_1}(\kappa^{-1}(a^*))},~~
\forall a \in A
\label{starchi}
\en
\noi where the overline denotes the usual complex conjugation.
As an example, let us consider the $q$-generators $\cchi{a}{b}$
in the fundamental
representation:
\eq
(\cchi{a_1}{a_2})^{b_1}_{~b_2} \equiv \cchi{a_1}{a_2} (\T{b_1}{b_2})
\en
\noi By using eqs.
(\ref{unitarity}) and (\ref{starchi}), it is easy to
find that
\eq
[(\cchi{a_1}{a_2})^*]^{b_1}_{~b_2}=
\overline{(\cchi{a_2}{a_1})^{b_2}_{~b_1}}
\en
\eq
\Rightarrow (\cchi{a_1}{a_2})^*=(\cchi{a_2}{a_1})^{\dag}
\label{starchi2}
\en
\noi where the dagger $\dag$ means usual hermitian conjugation
of the matrix that represents $\cchi{a_2}{a_1}$.
\sk

The $GL_q(3)$ Lie algebra relations are presented
in Table 4, with the standard index convention already
used for the $\T{a}{b}$
and the $\ome{a}{b}$. In the limit
$q \ra 1$, we recover the $SU(3) \otimes U(1)$ Lie algebra in
the Cartan basis. The
two commuting $U(1)$'s inside $SU(3)$ are generated by ${1 \over
{2\sqrt{3}}} (\chi_1-\chi_5)$
and ${1 \over 6}(\chi_1+\chi_5-2\chi_9)$, and the
remaining $U(1)$ by $\chi_1+
\chi_5+\chi_9$. The generators ${1 \over {\sqrt{6}}}
(\chi_2,\chi_3,\chi_6,\chi_4,\chi_7,\chi_8)$
correspond to the roots of the $SU(3)$ hexagon, ordered
counterclockwise, with
$\chi_2$ on the positive $x$-axis. Note that
the $q$-generators $\chi_5,\chi_9,
\chi_6,\chi_8$ close on the $GL_q(2)$ $q$-Lie
algebra discussed for ex. in
ref. \cite{Aschieri} (there the generators are
denoted by $\chi_1,\chi_2,\chi_+,\chi_-$).
In the limit $q \ra 1$, the rule (\ref{starchi2}) can
be verified to yield the familiar
relations $\chi_2=(\chi_4)^{\dag}$ etc.
\sk
The Cartan-Maurer equations are found by
applying to $\ome{c_1}{c_2}$ the exterior differential as defined in
 (\ref{exteriordifferential}):
\eq
d\ome{c_1}{c_2}=\lam (\ome{b}{b} \we \ome{c_1}{c_2} + \ome{c_1}{c_2} \we
\ome{b}{b}) \equiv -\cc{a_1}{a_2}{b_1}{b_2}{c_1}{c_2} ~\ome{a_1}{a_2} \we
\ome{b_1}{b_2}. \label{CartanMaurer}
\en
In order to obtain an explicit expression for the $C$ structure
constants in (\ref{CartanMaurer}) which reduces to the correct $q=1$
limit, we must use the relation
(\ref{commom}) for the commutations of $\ome{a_1}{a_2}$ with
$\ome{b_1}{b_2}$. Then the term $\ome{c_1}{c_2} \we \ome{b}{b}$
in (\ref{CartanMaurer}) can be written as
$-Z\om\om$ via formula (\ref{commom}), and we find the
$C$-structure constants to be:
\eqa
\cc{a_1}{a_2}{b_1}{b_2}{c_1}{c_2}&=&
-\lam(\de^{a_1}_{a_2} \de^{b_1}_{c_1}
\de^{c_2}_{b_2} - {1 \over {q^2 + q^{-2}}}[ \RRhat{c_1}{c_2}{b}{b}{a_1}
{a_2}{b_1}{b_2}+\RRhatinv{c_1}{c_2}{b}{b}{a_1}{a_2}{b_1}{b_2}])
\nonumber\\
&=& -\lam(\de^{a_1}_{a_2} \de^{b_1}_{c_1}
\de^{c_2}_{b_2} - {1 \over {q^2 + q^{-2}}}[ \de^{a_1}_{a_2}
\de^{b_1}_{c_1}\de^{c_2}_{b_2}+
\RRhatinv{c_1}{c_2}{b}{b}{a_1}{a_2}{b_1}{b_2}]),
\label{explicitcc}
\ena
\noi where we have used (see for example \cite{Aschieri})
\eq
\RRhat{c_1}{c_2}{b}{b}{a_1}{a_2}{b_1}{b_2} = \de^{a_1}_{a_2}
\de^{b_1}_{c_1} \de^{c_2}_{b_2}. \label{RReqdeltas}
\en
\noi By considering
the analogue of (\ref{RRffMM}) for $\Rh^{-1}$,
it is not difficult to see that the terms proportional to $\lam$
cancel, and the \qone limit of (\ref{explicitcc})
is well defined.

The Cartan-Maurer equations for $GL_q(3)$ are collected in Table 5.
\sk
Finally, we compute the Lie derivative of the $q$-group elements
$T_i$ and of the left-invariant 1-forms $\om^i$.
As discussed in \cite{Aschieri}, the functionals $\chi_i$ are
the $q$-analogue of tangent vectors at the origin of the group,
while $t_i=\chi * =(id \otimes\chi_i)\D$ are the q-analogue of left-
invariant vector fields.
The Lie derivative along $t=\chi *$ of a generic element $\tau \in
\Ga^{\otimes n}$ is defined by  \cite{Aschieri}
\eq
\ell_t(\tau)\equiv\chi * \tau \equiv (id \otimes \chi)\D_R(\tau)
\en
\noi We refer to \cite{Aschieri}
for a discussion of the properties of the
quantum Lie derivative.  The Lie
derivative of $\om^i$ is essential for defining the $q$-gauge
variations of the $q$-potentials $\om^i$, see
refs. \cite{Cas2} and \cite{Aschieri}.
Table 6 and Table 7 contain the
exterior derivative of the $\T{a}{b}$ elements of $GL_q(3)$,
and the Lie derivative of the
left-invariant $\ome{a}{b}$. The
relevant formulas are (\ref{qgenerators}) and:
\eq
\cchi{a_1}{a_2} * \T{b_1}{b_2}=
\lam \T{b_1}{d} [s\,\Rinv{da_1}{fe} \Rinv{ef}{a_2b_2}
-\de^{a_1}_{a_2} \de^{d}_{b_2}]
\en
\eq
\cchi{a_1}{a_2} * \ome{b_1}{b_2}=
\ome{c_1}{c_2} \CC{c_1}{c_2}{a_1}{a_2}{b_1}{b_2}
\en
\noi cf. ref. \cite{Aschieri}. In Table 7
we actually give the $q$-analogue of
the gauge variations of the
potentials $\om^i$, i.e. $\delta \om^i \equiv \epsi^k \ell_{t_k} \om^i$,
where $\epsi^k$ is a ``$q$-gauge parameter".\\
As in the classical case, the $q$-Lie
derivative commutes with the exterior differental:
\eq
d\:\ell_t=\ell_td
\en
\noi and is a representation
of the $q$-Lie algebra, since \cite{Aschieri}:
\eq
[\ell_{t_i},\ell_{t_j}]_q=\ell_{[t_i,t_j]_q}
\en
\noi with $[t_i,t_j]_q \equiv [\chi_i,\chi_j]_q~ *$. The $q$-commutator
$[\,~,~]_q$ is
defined by the left-hand side of eq. (\ref{qLiealgebra}).
\sk
We conclude by observing that the $GL_q(2)$ bicovariant differential
calculus of ref. \cite{Aschieri} can be obtained by ``projecting"
the $GL_q(3)$ formulas obtained in this Letter into their
$GL_q(2)$ subspace. This is done by rescaling the $T$, $\om$ and
$\chi$ with indices $1,2,3,4,7$ and letting the rescaling
parameter go to zero.

\vfill\eject
\centerline{\bf Table 1}
\centerline{The $R$ matrix for $GL_q(3)$}
$$
\R{ab}{cd}=\left(  \begin{array}{ccccccccc}
   q&0&0&0&0&0&0&0&0\\
   0&1&0&0&0&0&0&0&0\\
   0&0&1&0&0&0&0&0&0\\
   0&\lambda&0&1&0&0&0&0&0\\
0&0&0&0&q&0&0&0&0\\
0&0&0&0&0&1&0&0&0\\
0&0&\lambda&0&0&0&1&0&0\\
0&0&0&0&0&\lambda&0&1&0\\
0&0&0&0&0&0&0&0&q\\
\end{array} \right) \nonumber
$$
where $\lambda=q-q^{-1}~$ .
\sk\sk
\centerline{\bf Table 2}
\centerline{The commutation relations between the left-invariant $\om^i$}
$$
\om^1\om^1=\om^2\om^2=\om^3\om^3=\om^4\om^4=\om^6\om^6=\om^7\om^7=
\om^8\om^8=0
$$
$$
\om^1\om^2+\om^2\om^1=0
$$
$$
\om^1\om^3+\om^3\om^1=0
$$
$$
\om^1\om^4+\om^4\om^1=0
$$
$$
 \omega ^{1}\,\omega ^{5}+\omega ^{5}\,\omega ^{1}+\unmezzo (1-q^2)
   (\omega ^{4}\,\omega ^{2}-\omega ^{2}\,\omega ^{4})= 0
$$
$$
   \omega ^{1}\,\omega ^{6}+\omega ^{6}\,\omega ^{1}+\unmezzo (1-q^2)
   (\omega ^{4}\,\omega ^{3}-\qm\,\omega ^{3}\,\omega ^{4})= 0
$$
$$
\om^1\om^7+\om^7\om^1=0
$$
$$
   \omega ^{1}\,\omega ^{8}+\omega ^{8}\,\omega ^{1}+\unmezzo (1-q^2)
   (\omega ^{7}\,\omega ^{2}-\qm\,\omega ^{2}\,\omega ^{7}) = 0
$$
$$
   \omega ^{1}\,\omega ^{9}+\omega ^{9}\,\omega ^{1}+\unmezzo (1-q^2)
   (\omega ^{7}\,\omega ^{3}-\omega ^{3}\,
   \omega ^{7})+\unmezzo (1-q^2)^2\,\omega^2
   \,\omega^4 = 0
$$
$$
 \omega ^{2}\,\omega ^{3}+q\,\omega ^{3}\,\omega ^{2} = 0
$$
$$
(1+q^4)\,\omega ^{2}\,\omega ^{4}+2\,{q^2}\,\omega ^{4}\,\omega ^{2}+
(q^2-1)\,(\omega ^{1}\,\omega ^{5} + \omega ^{5}\,\omega ^{1}) = 0
$$
$$
{q^2}\,\omega ^{2}\,\omega ^{5}+\omega ^{5}\,\omega ^{2}
+(1-q^2)^2\,\omega ^{1}\,
\omega ^{2}+(1-q^2)\,\omega ^{2}\,\omega ^{1} = 0
$$
$$
\omega ^{2}\,\omega ^{6}+\qm \,
\omega ^{6}\,\omega ^{2}+\unmezzo (q^{-2}-1)\,
(\omega ^{5}\,\omega ^{3}-\omega ^{3}\,\omega ^{5})+
(q^2-1)\omega ^{1}\,\omega ^{3}
= 0
$$
$$
(1+q^4)\,\omega ^{2}\,
\omega ^{7} +2\,{q^3}\,\omega ^{7}\,\omega ^{2} +q(q^2-1)
(\omega ^{1}\,\omega ^{8}+\omega ^{8}\,\omega ^{1})=0
$$
$$
q\,\omega ^{2}\,\omega ^{8} +\omega ^{2}\,\omega ^{8} = 0
$$
$$
\omega ^{2}\,\omega ^{9}+\omega ^{9}\,\omega ^2+\unmezzo (q-\qm)
(q\,\omega ^{3}\,\omega ^{8}-
\omega ^{8}\,\omega ^{3})+\unmezzo q^2(1-q^2)^2\,
\omega ^{1}\,\omega ^{2}+\unmezzo (1-q^2)^2\,\omega ^{2}\,\omega ^{5}=0
$$
$$
(1+q^4)\,\omega ^{3}\,
\omega ^{4} +2\,{q^3}\,\omega ^{4}\,\omega ^{3} +q(q^2-1)
(\omega ^{1}\,\omega ^{6}+\omega ^{6}\,\omega ^{1})=0
$$
$$
(1+q^4)\,\omega ^{3}\,\omega ^{5} +
2\,{q^2}\,\omega ^{5}\,\omega ^{3} +q(q^2-1)
(\omega ^{2}\,\omega ^{6}+\qm\,\omega ^{6}\,\omega ^{2}) +
q^2(1-q^2)^2\,\om^1 \om^3=0
$$
$$
\om^3\om^6+\qm\,\om^6 \om^3=0
$$
$$
(1+q^4)\,\om^3\om^7+2q^2\,\om^7\om^3-(1-q^2)^2\,(\om^1\om^5+\om^5\om^1)
+(q^2-1) (\om^1\om^9+\om^9\om^1)+2q^2(1-q^2)\om^4\om^2=0
$$
$$
(1+q^4)\,\om^3\om^8+2q\,\om^8\om^3-(1-q^2)^2\,\om^2\om^5+
(1-q^4)\,\om^5\om^2
+(q^2-1) (\om^2\om^9+\om^9\om^2)-2q^2(1-q^2)\om^1\om^2=0
$$
$$
q^2(1+q^2)(\om^3\om^9+q^{-2}\,\om^9\om^3)-(1-q^2)(2-q^2+q^4)\,\om^1\om^3+
2q(1-q^2)\om^6\om^2+q^2(1-q^2)^2\om^2\om^6=0
$$
$$
(1+q^2)(\om^4\om^5+q^2\,\om^5\om^4)+q^2(q^4-1)\,\om^4\om^1=0
$$
$$
\om^4\om^6+q\,\om^6\om^4=0
$$
$$
\om^4\om^7+\qm\,\om^7\om^4=0
$$
$$
\om^4\om^8+q\om^8\om^4+\unmezzo \qm (q^2-1) (\om^5\om^7-\om^7\om^5)
+q(q^2-1)\om^7\om^1=0
$$
$$
\om^4\om^9+q\om^9\om^4+\unmezzo \qm (q^2-1) (\om^6\om^7-\qm\,\om^7\om^6)
=0
$$
$$
(1+q^2)^2\,\om^5\om^5 + (1-q^2)^2\,(\om^1\om^5+\om^5\om^1)+
2q^2(1-q^2)(\om^2\om^4-\om^4\om^2)=0
$$
$$
(1+q^2)(\om^5\om^6+\om^6\om^5)+(1-q^2)^2(\om^1\om^6+\om^6\om^1)+2q(1-q^2)
\om^3\om^4-q^2(q^2-1)^2\om^4\om^3=0
$$
$$(1+q^4)\,\om^5\om^7+2q^2\,\om^7\om^5+q^2(1-q^2)^2\om^7\om^1+
q^2(q^2-1)(\qm\,\om^4\om^8+\om^8\om^4)=0
$$
$$
\om^5\om^8+\om^8\om^5+(q^2-1) \,\om^7\om^2=0
$$
$$
\om^5\om^9+\om^9\om^5+\unmezzo q^2(q^2-1)^2\,\om^4\om^2-\unmezzo
(q^2-1)^2\,\om^7\om^3+\unmezzo (1-q^2)^2 \om^5\om^5+\unmezzo
(q^2-1)(\om^6\om^8-\om^8\om^6)=0
$$
$$
(1+q^4) \om^6\om^7+2q\om^7\om^6+
q^2(1-q^2)\,\om^1\om^4-(1-q^2)^2\om^5\om^4
+(1-q^4)\,\om^4\om^5+(q^2-1)(\om^4\om^9+\om^9\om^4)=0
$$
$$
(1+q^4)\,\om^6\om^8+2q^2\om^8\om^6+(1-q^2)^2(\om^1\om^5+\om^5\om^1)
-2q^2(1-q^2)^2 \,\om^4\om^2+2q^2(1-q^2)\om^2\om^4-
$$
$$
-2q^2(1-q^2)\,\om^7\om^3
+2q^2(1-q^2)\om^5\om^5+(q^2-1)(\om^5\om^9+\om^9\om^5)=0
$$
$$
q^2(q^2+1)(\om^6\om^9+q^{-2}\,\om^9\om^6)+
(q^2-1)^2\,(\om^1\om^6+\om^6\om^1)
+2q(1-q^2)\om^3\om^4+q^4(1-q^2)^2\,\om^4\om^3+
$$
$$
+(1-q^2-q^4+q^6)\om^5\om^6+
(1-q^4)\,\om^6\om^5=0
$$
$$
\om^7\om^8+q\,\om^8\om^7=0
$$
$$
(1+q^2)(\om^7\om^9+q^2\,\om^9\om^7)+q^4(q^4-1)\om^7\om^1+q^2(q^4-1)
\om^8\om^4=0
$$
$$
(1+q^2)(\om^8\om^9+q^2\,\om^9\om^8)+q^4(q^4-1)\om^7\om^2+q^2(q^4-1)
\om^8\om^5=0
$$
$$
(1+q^2)^2\,\om^9\om^9+(1-q^2)^2(\om^1\om^9+\om^9\om^1)-2q^2(1-q^2)^2
(\om^2\om^4+q^2\,\om^4\om^2)+2q^2(1-q^2)(\om^3\om^7-q^2\,\om^7\om^3)
$$
$$
-2q^2(1-q^2)^2\,\om^5\om^5+(1-q^2)^2\,(\om^5\om^9+\om^9\om^5)+
2q^2(q^2-1)(\om^8\om^6-\om^6\om^8)=0
$$
\vfill\eject
\centerline{\bf Table 3}
\centerline{The commutation relations between $\om^i$ and $T_j$}
\[
\begin{array}{llll}
\om^1T_1=q^{-2}T_1\om^1&\om^1T_2=T_2\om^1&\om^1T_3=T_3\om^1&
\om^1T_4=q^{-2}T_4\om^1\\
\om^1T_5=T_5\om^1&\om^1T_6=T_6\om^1&\om^1T_7=q^{-2}T_7\om^1&
\om^1T_8=T_8\om^1\\
\om^1T_9=T_9\om^1&\om^2T_1=q^{-1}T_1\om^2&\om^2T_3=T_3\om^2&
\om^2T_4=q^{-1}T_4\om^2\\
\om^2T_6=T_6\om^2&\om^2T_7=\qm T_7\om^2&\om^2T_9=T_9\om^2&
\om^3T_1=q^{-1}T_1\om^3\\
\om^3T_2=T_2\om^3&\om^3T_4=\qm T_4\om^3&\om^3T_5=T_5\om^3&
\om^3T_7=q^{-1}T_7\om^3\\
\om^3T_8=T_8\om^3&\om^4T_2=\qm T_2\om^4&\om^4T_3=T_3\om^4&
\om^4T_5=q^{-1}T_5\om^4\\
\om^4T_6=T_6\om^4&\om^4T_8=\qm T_8\om^4&\om^4T_9=T_9\om^4&
\om^5T_3=T_3\om^5\\
\om^5T_6=T_6\om^5&\om^5T_9=T_9\om^5&\om^6T_2=\qm T_2\om^6&
\om^6T_5=\qm T_5\om^6\\
\om^6T_8=q^{-1}T_8\om^6&\om^7T_3=\qm T_3\om^7&\om^7T_6=\qm T_7\om^6&
\om^7T_9=q^{-1}T_9\om^7\\
\om^8T_3=q^{-1}T_3\om^8&\om^8T_6=\qm T_6\om^8&\om^8T_9=\qm T_9\om^8&
\\
\end{array}
\]
\[
\begin{array}{ll}
\om^2T_2=(q^{-2}-1)T_1\om^1+\qm T_2\om^2&\om^2T_5=(q^{-2}-1)T_4\om^1+
\qm T_5\om^2\\
\om^2T_8=(q^{-2}-1)T_7\om^1+
\qm T_8\om^2&\om^4T_1=(q^{-2}-1)T_2\om^1+\qm T_1\om^4\\
\om^4T_4=(q^{-2}-1)T_5\om^1+\qm T_4\om^4&\om^4T_7=
(q^{-2}-1)T_8\om^1+\qm T_7\om^4\\
\om^5T_1=(q^{-1}-q)T_2\om^2+ T_1\om^5&\om^5T_4=(q^{-1}-q)T_5\om^2+
T_4\om^5\\
\om^5T_7=(q^{-1}-q)T_8\om^2+
T_7\om^5&\om^6T_1=(q^{-1}-q)T_2\om^3+ T_1\om^6\\
\om^6T_4=(q^{-1}-q)T_5\om^3+
T_4\om^6&\om^6T_7=(q^{-1}-q)T_8\om^3+ T_7\om^6\\
\om^7T_1=(q^{-2}-1)T_3\om^1+\qm T_1\om^7&\om^7T_2=(q^{-1}-q)T_3\om^4+
T_2\om^7\\
\om^7T_4=(q^{-2}-1)T_6\om^1
+\qm T_4\om^7&\om^7T_5=(q^{-1}-q)T_6\om^4+ T_5\om^7\\
\om^7T_7=(q^{-2}-1)T_9\om^1+
\qm T_7\om^7&\om^7T_8=(q^{-1}-q)T_9\om^4+ T_8\om^7\\
\om^8T_4=(q^{-1}-q)T_6\om^2+ T_4\om^8&\om^8T_7=(q^{-1}-q)T_9\om^2+
 T_7\om^8\\
\om^9T_1=(q^{-1}-q)T_3\om^3+
T_1\om^9&\om^9T_2=(q^{-1}-q)T_3\om^6+ T_2\om^9\\
\om^9T_4=(q^{-1}-q)T_6\om^3+T_4\om^9&\om^9T_5=
(q^{-1}-q)T_6\om^6+ T_5\om^9\\
\om^9T_7=(q^{-1}-q)T_9\om^3+ T_7\om^9&\om^9T_8=(q^{-1}-q)T_9\om^6+
T_8\om^9\\
\end{array}
\]
\[
\begin{array}{l}
\om^3T_3=(q^{-2}-1)T_1\om^1+(\qm-q)T_2\om^2+\qm T_3\om^3\\
\om^3T_6=(q^{-2}-1)T_4\om^1+(\qm-q)T_5\om^2+\qm T_6\om^3\\
\om^3T_9=(q^{-2}-1)T_7\om^1+(\qm-q)T_8\om^2+\qm T_9\om^3\\
\om^5T_2=(q^{-1}-q)^2\,T_2\om^1+(\qm-q)T_1\om^4+q^{-2}T_2\om^5\\
\om^5T_5=(q^{-1}-q)^2\,T_5\om^1+(\qm-q)T_4\om^4+q^{-2}T_5\om^5\\
\om^5T_8=(q^{-1}-q)^2\,T_8\om^1+(\qm-q)T_7\om^4+q^{-2}T_8\om^5\\
\om^6T_3=(\qm-q)^2T_2\om^1+(\qm-q)T_1\om^4+(q^{-2}-1)T_2\om^5+
\qm T_3\om^6\\
\om^6T_6=(\qm-q)^2T_5\om^1+(\qm-q)T_4\om^4+(q^{-2}-1)T_5\om^5+
\qm T_6\om^6\\
\om^6T_9=(\qm-q)^2T_8\om^1+(\qm-q)T_7\om^4+(q^{-2}-1)T_8\om^5+
\qm T_9\om^6\\
\om^8T_2=(\qm-q)^2T_3\om^1+(\qm-q)T_1\om^7+(q^{-2}-1)T_3\om^5+
\qm T_2\om^8\\
\om^8T_5=(\qm-q)^2T_6\om^1+(\qm-q)T_4\om^7+(q^{-2}-1)T_6\om^5+
\qm T_5\om^8\\
\om^8T_8=(\qm-q)^2T_9\om^1+(\qm-q)T_7\om^7+(q^{-2}-1)T_9\om^5+
\qm T_8\om^8\\
\om^9T_3=(\qm-q)^2T_3\om^1+(\qm-q)^2T_3\om^5+(\qm-q)T_1\om^7+
(\qm-q)T_2\om^8+q^{-2}T_3\om^9\\
\om^9T_6=(\qm-q)^2T_6\om^1+(\qm-q)^2T_6\om^5+(\qm-q)T_4\om^7+
(\qm-q)T_5\om^8+q^{-2}T_6\om^9\\
\om^9T_9=(\qm-q)^2T_9\om^1+(\qm-q)^2T_9\om^5+(\qm-q)T_7\om^7+
(\qm-q)T_8\om^8+q^{-2}T_9\om^9\\
\end{array}
\]
\sk
\vfill\eject
\centerline{\bf Table 4}
\centerline{The $q$-Lie algebra}
\def\1{{\chi_1}}
\def\2{{\chi_2}}
\def\3{{\chi_3}}
\def\4{{\chi_4}}
\def\5{{\chi_5}}
\def\6{{\chi_6}}
\def\7{{\chi_7}}
\def\8{{\chi_8}}
\def\9{{\chi_9}}
\[
\begin{array}{l}
\1\2-\2\1+(1-q^2)(\2\5+\3\8)=q\2\\[.1em]
\1\3-\3\1+(1-q^2)(q^2\9\3+\2\6)=q^3\3\\[.1em]
\1\4-\4\1-(1-q^2)(\5\4+\6\7)=-q\4\\[.1em]
\1\5-\5\1=0\\[.1em]
\1\6-\6\1=0\\[.1em]
\1\7-\7\1-(1-q^2)(\8\4+\9\7)=-q\7\\[.1em]
\1\8-\8\1=0\\[.1em]
\1\9-\9\1=0\\[.1em]
\2\3-q^{-1}\3\2=0\\[.1em]
\2\4-\4\2+(1-q^2)[\5\1-\5\5+
\7\3-q^2\8\6-(1-q^2)(\9\1-\9\5)]~~~~\\[-0.1em]
{}~~~~~~~~~~~~~~~~~~~~~~~~~~~~~~~~~~~~~~~~~~~~~~~~~~~
{}~~~~~~~~~~~~~~~~~~~~~~~~~~~~~~~~~~~~~~~~~=q^3(\1-\5)\\[-0.45em]
\2\5-q^2\5\2+(1-q^2)\3\8=q\2\\[.1em]
\2\6-q\6\2+(1-q^2)\3\9=q\3\\[.1em]
\2\7-q^{-1}\7\2+(q^{-1}-q)(\8\1-\8\5-\9\8)=-\8\\[.1em]
\2\8-q\8\2=0\\[.1em]
\2\9-\9\2=0\\[.1em]
\3\4-q^{-1}\4\3+(q^{-1}-q)(\6\1-\5\6-\6\9)=-\6\\[.1em]
\3\5-\5\3+(q^{-1}-q)\6\2=0\\[.1em]
\3\6-q\6\3=0\\[.1em]
\3\7-\7\3-(1-q^2)(\8\6-\9\1+\9\9)=q(\1-\9)\\[.1em]
\3\8-q\8\3+(1-q^2)\9\2=q\2\\[.1em]
\3\9-q^2\9\3=q\3\\[.1em]
\4\5-q^{-2}\5\4+(1-q^{-2})\6\7=-q^{-1}\4\\[.1em]
\4\6-q^{-1}\6\4=0\\[.1em]
\4\7-q\7\4=0\\[.1em]
\4\8-q^{-1}\8\4-(q^{-1}-q)\9\7=-\7\\[.1em]
\4\9-\9\4=0\\[.1em]
\5\6-\6\5+(1-q^2)\6\9=q\6\\[.1em]
\5\7-\7\5+(q^{-1}-q)\4\8=0\\[.1em]
\5\8-\8\5-(1-q^2)\9\8=-q\8\\[.1em]
\5\9-\9\5=0\\[.1em]
\6\7-q\7\6+(1-q^2)\9\4=q\4\\[.1em]
\6\8-\8\6+(1-q^2)(\9\5-\9\9)=q(\5-\9)\\[.1em]
\6\9-q^2\9\6=q\6\\[.1em]
\7\8-q^{-1}\8\7=0\\[.1em]
\7\9-q^{-2}\9\7=q^{-1}\7\\[.1em]
\8\9-q^{-2}\9\8=-q^{-1}\8
\end{array}
\]
\vfill\eject
\centerline{\bf Table 5}
\centerline{The Cartan-Maurer equations}
\[
\begin{array}{ll}
(1+q^4)\,d\omega ^{1}=& - q\,( 1 - {q^2} ) \,
       \omega ^{1}   \omega ^{5} -
   q\,( 1 - {q^2} ) \,\omega ^{1}   \omega ^{9}
    - {q^5}\omega ^{2}   \omega ^{4}
   - {q^3}\omega ^{3}   \omega ^{7}\\
& + {q^3}\omega ^{4}   \omega ^{2}
     - q( 1 - {q^2} ) \,
       \omega ^{5}   \omega ^{1}
       +  {q^3}\,\omega ^{7}   \omega ^{3}-
   q\,( 1 - {q^2} ) \,\omega ^{9}   \omega ^{1}\\
(1+q^4)\,d\omega ^{2}=&-{q^7}\,\omega ^{1}   \omega ^{2}
+ {q^3}\,\omega ^{2}   \omega ^{1}- q\,( 1 - {q^2} + {q^4} ) \,
       \omega ^{2}   \omega ^{5} -
   q\,( 1 - {q^2} ) \,\omega ^{2}   \omega ^{9}\\
   &- {q^3}\,\omega ^{3}   \omega ^{8} + {q^3}\,\omega ^{5}   \omega ^{2}
     + {q^2}\,\omega ^{8}   \omega ^{3}
     - q\,( 1 - {q^2} ) \,
       \omega ^{9}   \omega ^{2}\\
(1+q^4)\,d\omega ^{3}=&-{q^7}\,\omega ^{1}   \omega ^{3}
- {q^5}\,\omega ^{2}   \omega ^{6} + {q^3}\,\omega ^{3}   \omega ^{1}
- q\,( 1 - {q^2} ) \,
       \omega ^{3}   \omega ^{5} -
   q\,\omega ^{3}   \omega ^{9} +
   {q^2}\,\omega ^{6}   \omega ^{2} +
   {q^3}\,\omega ^{9}   \omega ^{3}\\
(1+q^4)\,d\omega ^{4}=&{q^3}\,\omega ^{1}   \omega ^{4} -
   {q^7}\,\omega ^{4}   \omega ^{1}+
   {q^3}\,\omega ^{4}   \omega ^{5} -
   q\,( 1 - {q^2} ) \,\omega ^{4}   \omega ^{9}\\
   &- q\,( 1 - {q^2} + {q^4} ) \,
       \omega ^{5}   \omega ^{4}-
   {q^3}\,\omega ^{6}   \omega ^{7} +
   {q^2}\,\omega ^{7}   \omega ^{6}-
   q\,( 1 - {q^2} ) \,\omega ^{9}   \omega ^{4}\\
(1+q^4)\,d\omega ^{5}=& {q^3}\,\omega ^{2}   \omega ^{4} -
   {q^3}\,( 1 - {q^2} + {q^4} ) \,
       \omega ^{4}   \omega ^{2} -
   q\,{{( -1 + {q^2} ) }^2}\,\omega ^{5}   \omega ^{5}
  - q\,( 1 - {q^2} ) \,
       \omega ^{5}   \omega ^{9}\\
  &- {q^3}\,\omega ^{6}   \omega ^{8} -
   {q^3}\,( 1 - {q^2} ) \,\omega ^{7}   \omega ^{3}
   + {q^3}\,\omega ^{8}   \omega ^{6}- q\,( 1 - {q^2} ) \,
       \omega ^{9}   \omega ^{5}\\
(1+q^4)\,d\omega ^{6}=&{q^2}\,\omega ^{3}   \omega ^{4}-
   {q^7}\,\omega ^{4}   \omega ^{3} -
   {q^5}\,\omega ^{5}   \omega ^{6} +
   {q^3}\,\omega ^{6}   \omega ^{5}-
   q\,\omega ^{6}   \omega ^{9}+
   {q^3}\,\omega ^{9}   \omega ^{6}\\
(1+q^4)\,d\omega ^{7}=&{q^3}\,\omega ^{1}   \omega ^{7} +
   {q^2}\,\omega ^{4}   \omega ^{8} -
   q\,( 1 - {q^2} ) \,\omega ^{5}   \omega ^{7}
   - {q^7}\,\omega ^{7}   \omega ^{1}\\
  &+ {q^3}\,\omega ^{7}   \omega ^{9}
     - {q^5}\,\omega ^{8}   \omega ^{4}
     - q\,\omega ^{9}   \omega ^{7}\\
(1+q^4)\,d\omega ^{8}=&{q^2}\,\omega ^{2}   \omega ^{7} +
   {q^3}\,\omega ^{5}   \omega ^{8}-
   {q^7}\,\omega ^{7}   \omega ^{2} -
   {q^5}\,\omega ^{8}   \omega ^{5} +
   {q^3}\,\omega ^{8}   \omega ^{9} -
   q\,\omega ^{9}   \omega ^{8}\\
(1+q^4)\,d\omega ^{9}=&-{q^3}\,( 1 - {q^2} ) \,\omega ^{2}   \omega ^{4}
   + {q^3}\,\omega ^{3}   \omega ^{7}
     - {q^5}\,( 1 - {q^2} ) \,
       \omega ^{4}   \omega ^{2}
  - {q^3}\,( 1 - {q^2} ) \,\omega ^{5}   \omega ^{5}\\
      & + {q^3}\,\omega ^{6}   \omega ^{8}
     - {q^5}\,\omega ^{7}   \omega ^{3}
     - {q^3}\,\omega ^{8}   \omega ^{6}- q\,( 1 - {q^2} ) \,
       \omega ^{9}   \omega ^{9}
\end{array}
\]
\sk
\centerline{\bf Table 6}
\centerline{The exterior derivative of $T_i$}
$$
d\,T_{1} = {{( -{q^2} + s ) }\over
{-q + {q^3}}}\,T_{1}\,\omega ^{1} - s\,T_{2}\,
\omega ^{2} - s\,T_{3}\,\omega ^{3} +
    {{( -1 + s ) }\over {-\qm + q}}\,T_{1}\,(\omega ^{5} +
    \omega^{9})
$$
$$
d\,T_{2} = {{( -{q^2} + s -
{q^2}\,s + {q^4}\,s )}\over {-q + {q^3}}} \,T_{2}\,
\omega ^{1} - s\,T_{1}\,\omega ^{4} +
    {{( -{q^2} + s )}\over {-q + {q^3}}} \,T_{2}\,\omega ^{5} -
    s\,T_{3}\,\omega ^{6} + {{( -1 + s ) }\over
        {-\qm + q}} \,T_{2}\,\omega ^{9}
$$
$$
d\,T_{3} =  {{( -{q^2} + s - {q^2}\,s + {q^4}\,s ) }\over {-q + {q^3}}}
\,T_{3}\,(\omega ^{1} +\om^5)
 - s\,T_{1}\,\omega ^{7} -
    s\,T_{2}\,\omega ^{8} + {{( -{q^2} + s ) }\over {-q + {q^3}}}
    \,T_{3}\,\omega ^{9}
$$
$$
d\,T_{4} = {{( -{q^2} + s ) }\over
      {-q + {q^3}}}\,T_{4}\,\omega^1 - s\,T_{5}\,
\omega ^{2} - s\,T_{6}\,\omega ^{3} +
    {{( -1 + s ) }\over {-\qm + q}} \,T_{4}\,(\omega ^{5}+\om^9)
$$
$$
d\,T_{5} =  {{( -{q^2} + s - {q^2}\,s + {q^4}\,s )}
\over {-q + {q^3}}} \,T_{5}\, \omega ^{1}- s\,T_{4}\,\omega ^{4} +
    {{( -{q^2} + s ) }\over {-q + {q^3}}}\,T_{5}\,\omega ^{5} -
    s\,T_{6}\,\omega ^{6} + {{( -1 + s )}\over
        {-\qm + q}} \,T_{5}\,\omega ^{9}
$$
$$
d\,T_{6} =  {{( -{q^2} + s - {q^2}\,s + {q^4}\,s ) }\over {-q + {q^3}}}
\,T_{6}\,(\omega ^{1}+\om^5)
- s\,T_{4}\,\omega ^{7} -
    s\,T_{5}\,\omega ^{8} +
{{( -{q^2} + s ) }\over {-q + {q^3}}} \,T_{6}\,\omega ^{9}
$$
$$
d\,T_{7} =  {{( -{q^2} + s ) }\over
      {-q + {q^3}}}\,T_{7}\,
\omega ^{1} - s\,T_{8}\,\omega ^{2} - s\,T_{9}\,\omega ^{3} +
    {{( -1 + s ) }\over {-\qm + q}} \,T_{7}\,(\omega ^{5}+\om^9)
$$
$$
d\,T_{8} =  {{( -{q^2} + s -
{q^2}\,s + {q^4}\,s ) }\over {-q + {q^3}}} \,T_{8}\, \omega ^{1}
- s\,T_{7}\,\omega ^{4} +
    {{( -{q^2} + s ) }\over {-q + {q^3}}}\,T_{8}\,\omega ^{5} -
    s\,T_{9}\,\omega ^{6} + {{( -1 + s ) }\over
        {-\qm + q}}\,T_{8}\,\omega ^{9}
$$
$$
d\,T_{9} =  {{( -{q^2} + s - {q^2}\,s + {q^4}\,s ) }\over {-q + {q^3}}}
\,T_{9}\,(\omega ^{1}+\om^5)
 - s\,T_{7}\,\omega ^{7} -
    s\,T_{8}\,\omega ^{8} +
{{( -{q^2} + s ) }\over {-q + {q^3}}} \,T_{9}\,\omega ^{9}
$$
\sk
\centerline{\bf Table 7}
\centerline{The Lie derivative of $\om^i$}
$$
\epsilon ^{i}\,{\ell}_{t_{i}}\,\omega ^{1} =
  {q^3}\,\epsilon ^{4}\,\omega ^{2} + q\,\epsilon ^{7}\,\omega ^{3} -
   q\,\epsilon ^{2}\,\omega ^{4} - q\,\epsilon ^{3}\,\omega ^{7} +
   \epsilon ^{1}\,[ ( -q + {q^5} ) \,\omega ^{1} +
      ( \qm - q ) \,\omega ^{5} +
      ( \qm - q ) \,\omega ^{9} ]
$$
$$
\epsilon ^{i}\,{\ell}_{t_{i}}\,\omega ^{2} =
  -\qm\,\epsilon ^{1}\,\omega ^{2} +
   {q^3}\,\epsilon ^{5}\,\omega ^{2} + q\,\epsilon ^{8}\,\omega ^{3} -
   \epsilon ^{3}\,\omega ^{8} +
   \epsilon ^{2}\,[ {q^5}\,\omega ^{1} - q\,\omega ^{5} +
      ( \qm - q ) \,\omega ^{9} ]
$$
$$
\epsilon ^{i}\,{\ell}_{t_{i}}\,\omega ^{3} =
  {q^3}\,\epsilon ^{6}\,\omega ^{2} - \qm\,\epsilon ^{1}\,\omega ^{3} +
   q\,\epsilon ^{9}\,\omega ^{3} - \epsilon ^{2}\,\omega ^{6} +
   \epsilon ^{3}\,( {q^5}\,\omega ^{1} - q\,\omega ^{9} )
$$
$$
\epsilon ^{i}\,{\ell}_{t_{i}}\,\omega ^{4} =
  ( -\qm + q + {q^5} ) \,\epsilon ^{1}\,\omega ^{4} -
   \qm\,\epsilon ^{5}\,\omega ^{4} + q\,\epsilon ^{7}\,\omega ^{6} -
   \epsilon ^{6}\,\omega ^{7} +
   \epsilon ^{4}\,[ - q\,\omega ^{1}  +
      ( \qm - q + {q^3} ) \,\omega ^{5} +
      ( \qm - q ) \,\omega ^{9} ]
$$
$$
\epsilon ^{i}\,{\ell}_{t_{i}}\,\omega ^{5} =
  - q\,\epsilon ^{4}\,\omega ^{2}   +
   ( q - {q^3} + {q^5} ) \,\epsilon ^{2}\,\omega ^{4} +
   \epsilon ^{1}\,( ( q - {q^3} ) \,\omega ^{1} +
      ( -\qm + q ) \,\omega ^{5} )  +
   q\,\epsilon ^{8}\,\omega ^{6} +
$$
$$
   ( q - {q^3} ) \,\epsilon ^{3}\,\omega ^{7} -
   q\,\epsilon ^{6}\,\omega ^{8} +
   \epsilon ^{5}\,[ ( -q + {q^3} ) \,\omega ^{5} +
      ( \qm - q ) \,\omega ^{9} ]
$$
$$
\epsilon ^{i}\,{\ell}_{t_{i}}\,\omega ^{6} =
  -\epsilon ^{4}\,\omega ^{3}   +
   {q^5}\,\epsilon ^{3}\,\omega ^{4} +
   ( -\qm + q ) \,\epsilon ^{1}\,\omega ^{6} -
   \qm\,\epsilon ^{5}\,\omega ^{6} + q\,\epsilon ^{9}\,\omega ^{6} +
   \epsilon ^{6}\,( {q^3}\,\omega ^{5} - q\,\omega ^{9} )
$$
$$
\epsilon ^{i}\,{\ell}_{t_{i}}\,\omega ^{7} =
  - \epsilon ^{8}\,\omega ^{4}   +
   ( -\qm + q + {q^5} ) \,\epsilon ^{1}\,\omega ^{7} +
   ( -\qm + q ) \,\epsilon ^{5}\,\omega ^{7} -
   \qm\,\epsilon ^{9}\,\omega ^{7} +
$$
$$
   {q^3}\,\epsilon ^{4}\,\omega ^{8} +
   \epsilon ^{7}\,[ -q\,\omega ^{1}  +
      ( \qm - q ) \,\omega ^{5} + \qm\,\omega ^{9} ]
$$
$$
\epsilon ^{i}\,{\ell}_{t_{i}}\,\omega ^{8} =
  - \epsilon ^{7}\,\omega ^{2}  +
   {q^5}\,\epsilon ^{2}\,\omega ^{7} +
   ( -\qm + q ) \,\epsilon ^{1}\,\omega ^{8} +
   ( -\qm + q + {q^3} ) \,\epsilon ^{5}\,\omega ^{8} -
   \qm\,\epsilon ^{9}\,\omega ^{8} +
   \epsilon ^{8}\,( - q\,\omega ^{5}  +
      \qm\,\omega ^{9})
$$
$$
\epsilon ^{i}\,{\ell}_{t_{i}}\,\omega ^{9} =
  ( q - {q^3} ) \,\epsilon ^{4}\,\omega ^{2} -
   q\,\epsilon ^{7}\,\omega ^{3} +
   ( {q^3} - {q^5} ) \,\epsilon ^{2}\,\omega ^{4} -
   q\,\epsilon ^{8}\,\omega ^{6} + {q^3}\,\epsilon ^{3}\,\omega ^{7} +
   q\,\epsilon ^{6}\,\omega ^{8} +
$$
$$
   \epsilon ^{1}\,[ ( {q^3} - {q^5} ) \,\omega ^{1} +
      ( -\qm + q ) \,\omega ^{9} ]  +
   \epsilon ^{5}\,[( q - {q^3} ) \,\omega ^{5} +
      ( -\qm + q ) \,\omega ^{9} ]
$$
\vfill\eject

\vfill\eject
\end{document}